\pdfoutput=1
\documentclass[runningheads]{llncs}
\usepackage[T1]{fontenc}
\usepackage{graphicx}
\usepackage{algorithm}
\usepackage{listings}
\usepackage{amsmath} 
\usepackage{amssymb}
\usepackage[table,dvipnames]{xcolor}
\usepackage{svg}
\usepackage{multirow}
\usepackage{adjustbox}
\usepackage{subcaption}
\usepackage{hyperref}

\definecolor{tablegray}{rgb}{0.92,0.92,0.92}
\definecolor{color1}{HTML}{792500}
\definecolor{color2}{HTML}{473992}

\newcommand{\Colorone}[1]{\textbf{#1}} 
\newcommand{\Colortwo}[1]{\underline{#1}}

\definecolor{codegreen}{rgb}{0,0.6,0}
\definecolor{codegray}{rgb}{0.5,0.5,0.5}
\definecolor{codepurple}{rgb}{0.58,0,0.82}
\definecolor{backcolour}{rgb}{0.95,0.95,0.92}

\lstdefinestyle{mystyle}{
    backgroundcolor=\color{backcolour},   
    commentstyle=\color{codegreen},
    keywordstyle=\color{magenta},
    numberstyle=\tiny\color{codegray},
    stringstyle=\color{codepurple},
    basicstyle=\ttfamily\footnotesize,
    breakatwhitespace=false,         
    breaklines=true,                 
    captionpos=b,                    
    keepspaces=true,                 
    numbers=left,                    
    numbersep=5pt,                  
    showspaces=false,                
    showstringspaces=false,
    showtabs=false,                  
    tabsize=2
}

\begin{document}
\title{MedNeXt: Transformer-driven Scaling of ConvNets for Medical Image Segmentation}
%
\titlerunning{MedNeXt: Transformer-driven Scaling of ConvNets}
%
\authorrunning{Roy et al.}
%
\author{Saikat Roy\inst{1,3} \and Gregor Koehler \inst{1} \and Constantin Ulrich \inst{1,5} \and Michael Baumgartner \inst{1,3,4} \and Jens Petersen \inst{1} \and Fabian Isensee \inst{1,4} \and Paul F. Jaeger \inst{4,6} \and Klaus Maier-Hein \inst{1,2}}

\institute{
Division of Medical Image Computing (MIC), German Cancer Research Center (DKFZ), Heidelberg, Germany \and
Pattern Analysis and Learning Group, Department of Radiation Oncology, Heidelberg University Hospital \and
Faculty of Mathematics and Computer Science, Heidelberg University, Germany \and
Helmholtz Imaging, German Cancer Research Center, Heidelberg, Germany \and
National Center for Tumor Diseases (NCT), NCT Heidelberg, A partnership between DKFZ and University Medical Center Heidelberg \and
Interactive Machine Learning Group, German Cancer Research Center, Heidelberg, Germany
}
\maketitle              
\begin{abstract}
There has been exploding interest in embracing Transformer-based architectures for medical image segmentation. However, the lack of large-scale annotated medical datasets make achieving performances equivalent to those in natural images challenging. Convolutional networks, in contrast, have higher inductive biases and consequently, are easily trainable to high performance. Recently, the ConvNeXt architecture attempted to modernize the standard ConvNet by mirroring Transformer blocks. In this work, we improve upon this to design a modernized and scalable convolutional architecture customized to challenges of data-scarce medical settings. We introduce MedNeXt, a \textit{Transformer-inspired} large kernel segmentation network which introduces -- 1) A \textit{fully} ConvNeXt 3D Encoder-Decoder Network for medical image segmentation, 2) Residual ConvNeXt up and downsampling blocks to preserve semantic richness across scales, 3) A novel technique to iteratively increase kernel sizes by upsampling small kernel networks, to prevent performance saturation 
on limited medical data, 4) Compound scaling at multiple levels (depth, width, kernel size) of MedNeXt. This leads to state-of-the-art performance on 4 tasks on CT and MRI modalities and varying dataset sizes, representing a \textit{modernized} deep architecture for medical image segmentation. Our code is made publicly available at: \href{https://github.com/MIC-DKFZ/MedNeXt}{https://github.com/MIC-DKFZ/MedNeXt}.

\end{abstract}

\section{Introduction}

Transformers \cite{vaswani2017attention,dosovitskiy2020image,liu2021swin} have seen wide-scale adoption in medical image segmentation as either components of hybrid architectures \cite{chen2021transunet,hatamizadeh2022unetr,xie2021cotr,cao2021swin,hatamizadeh2022swin,wang2021transbts} or standalone techniques \cite{zhou2021nnformer,peiris2022robust,karimi2021convolution} for state-of-the-art performance. The ability to learn long-range spatial dependencies is one of the major advantages of the Transformer architecture in visual tasks. However, Transformers are plagued by the necessity of large annotated datasets to maximize performance benefits owing to their limited inductive bias. While such datasets are common to natural images (ImageNet-1k \cite{deng2009imagenet}, ImageNet-21k \cite{ridnik2021imagenet}), medical image datasets usually suffer from the lack of abundant high quality annotations \cite{litjens2017survey}. To retain the inherent inductive bias of convolutions while taking advantage of architectural improvements of Transformers, the ConvNeXt \cite{liu2022convnet} was recently introduced to re-establish the competitive performance of convolutional networks for natural images. The ConvNeXt architecture uses an inverted bottleneck mirroring that of Transformers, composed of a depthwise layer, an expansion layer and a contraction layer (Sec. \ref{sec:network}), in addition to large depthwise kernels to replicate their scalability and long-range representation learning. The authors paired large kernel ConvNeXt networks with enormous datasets to outperform erstwhile state-of-the-art Transformer-based networks.
In contrast, the VGGNet \cite{simonyan2014very} approach of stacking small kernels continues to be the predominant technique for designing ConvNets in medical image segmentation. Out-of-the-box data-efficient solutions such as nnUNet \cite{isensee2021nnu}, using variants of a standard UNet \cite{cciccek20163d}, have still remained effective across a wide range of tasks. 

The ConvNeXt architecture marries the scalability and long-range spatial representation learning capabilities of Vision \cite{dosovitskiy2020image} and Swin Transformers \cite{liu2021swin} with the inherent inductive bias of ConvNets. Additionally, the inverted bottleneck design allows us to scale width (increase channels) while not being affected by kernel sizes. Effective usage in medical image segmentation would allow benefits from -- \textbf{1)} learning long-range spatial dependencies via large kernels, \textbf{2)} less intuitively, simultaneously scaling multiple network levels. To achieve this would require techniques to combat the tendency of large networks to overfit on limited training data. Despite this, there have been recent attempts to introduce large kernel techniques to the medical vision domain. In \cite{li2023large}, a large kernel 3D-UNet \cite{cciccek20163d} was used by decomposing the kernel into depthwise and depthwise dilated kernels for improved performance in organ and brain tumor segmentation -- exploring kernel scaling, while using constant number of layers and channels. The ConvNeXt architecture itself was utilized in 3D-UX-Net \cite{lee20223d}, where the Transformer of SwinUNETR \cite{hatamizadeh2022swin} was replaced with ConvNeXt blocks for high performance on multiple segmentation tasks. However, 3D-UX-Net only uses these blocks partially in a standard convolutional encoder, limiting their possible benefits.

In this work, we maximize the potential of a ConvNeXt design while uniquely addressing challenges of limited datasets in medical image segmentation. We present the first \textit{fully} ConvNeXt 3D segmentation network, \textbf{MedNeXt}, which is a scalable Encoder-Decoder network, and make the following contributions:

\begin{itemize}
    \item We utilize an architecture composed \textbf{purely of ConvNeXt blocks} which enables network-wide advantages of the ConvNeXt design. (Sec. \ref{sec:network})
    \item We introduce \textbf{Residual Inverted Bottlenecks} in place of regular up and downsampling blocks, to preserve contextual richness while resampling to benefit dense segmentation tasks. The modified residual connection in particular improves gradient flow during training.  (Sec.
    \ref{sec:resampling}) 
    \item We introduce a simple but effective technique of iteratively increasing kernel size, \textbf{UpKern}, to prevent performance saturation on large kernel MedNeXts by initializing with trained upsampled small kernel networks. (Sec. \ref{sec:upkern})
    \item We propose applying \textbf{Compound Scaling} \cite{tan2019efficientnet} of multiple network parameters owing to our network design, allowing orthogonality of width (\textit{channels}), receptive field (\textit{kernel size}) and depth (\textit{number of layers}) scaling. (Sec. \ref{sec:scaling})
\end{itemize}

MedNeXt achieves state-of-the-art performance against baselines consisting of Transformer-based, convolutional and large kernel networks. We show performance benefits on 4 tasks of varying modality (CT, MRI) and sizes (ranging from 30 to 1251 samples), encompassing segmentation of organs and tumors. We propose MedNeXt as a strong and modernized alternative to standard ConvNets for building deep networks for medical image segmentation.

\section{Proposed Method}

\begin{figure}[t]
    \centering
    \begin{subfigure}[b]{\textwidth}
    \includegraphics[trim={0 0.62cm 0 0},clip, width=\textwidth]{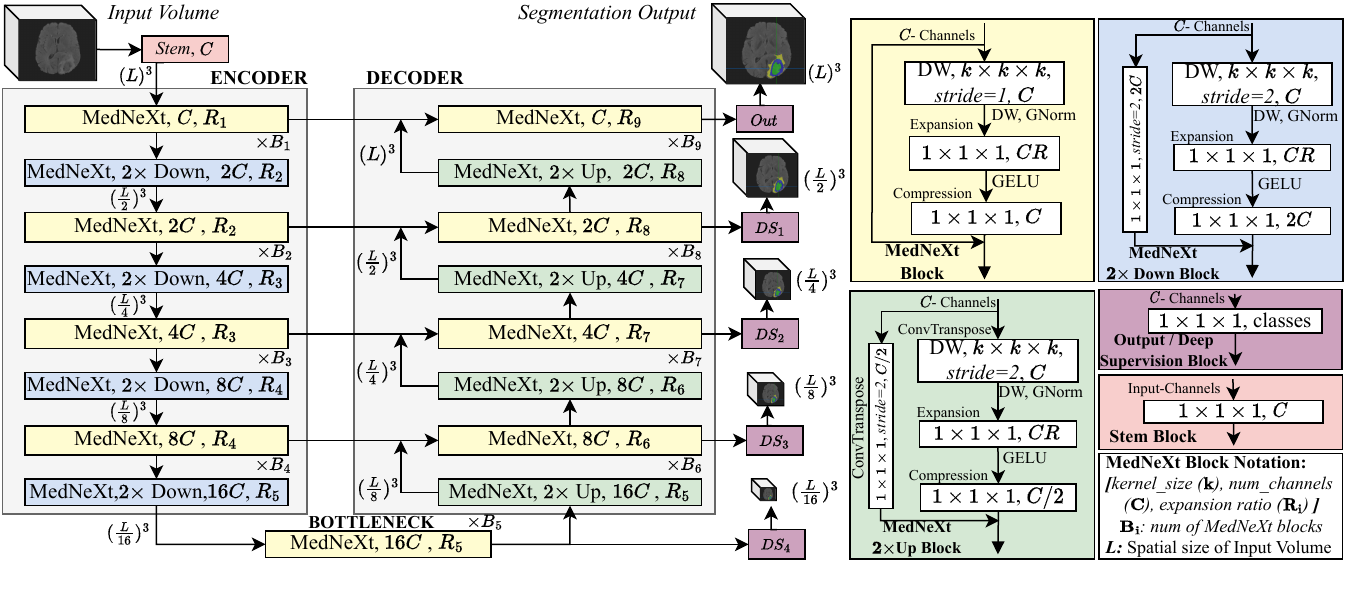}
    \caption{\textbf{MedNeXt macro and block architecture}}
    \label{fig:network_design}
    \end{subfigure}
    \begin{subfigure}[b]{0.55\textwidth}
    \centering
    \includegraphics[trim={0 0.2cm 0.1cm, 0},clip,width=\textwidth, height=3.0cm]{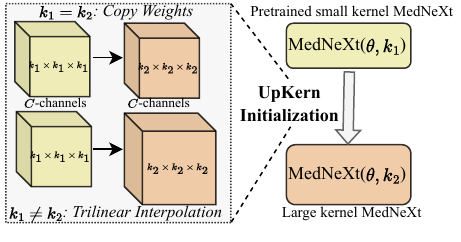}
    \caption{\textbf{UpKern Initialization}}
    \label{fig:upkern}
    \end{subfigure}
    \begin{subfigure}[b]{0.44\textwidth}
    \centering
    \includegraphics[trim={0.2cm 0.4cm 0.2cm 0.2cm},clip,width=0.97\textwidth, height=3.0cm]{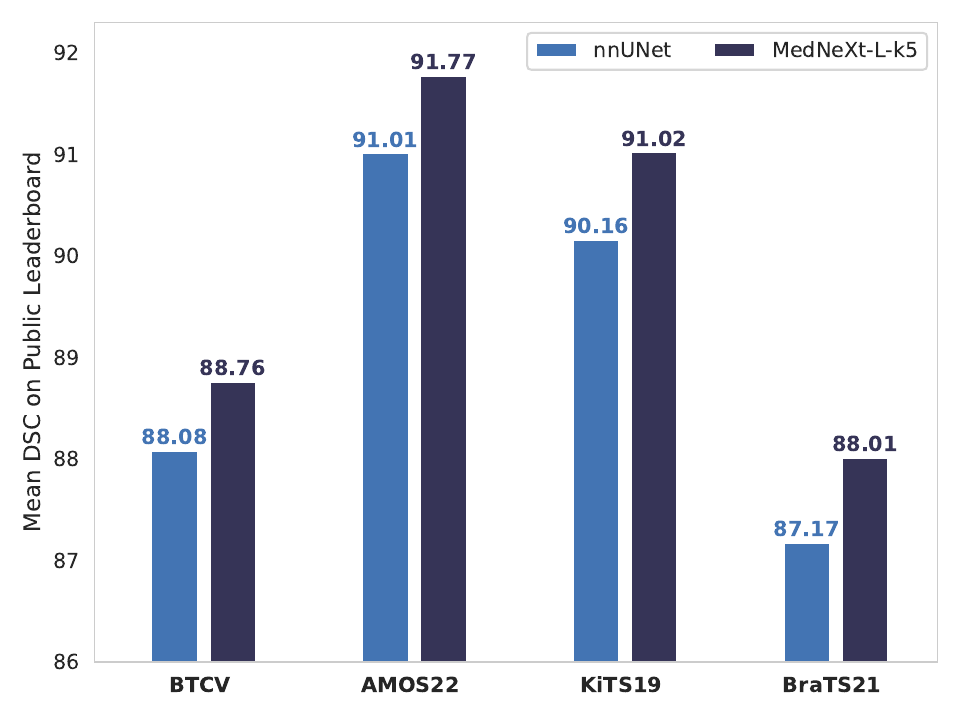}
    \caption{\textbf{Leaderboard Scores}}
    \label{fig:testset_scores}
    \end{subfigure}
    \caption{\textbf{(a)} Architectural design of the MedNeXt. The network has 4 Encoder and Decoder layers each, with a bottleneck layer. MedNeXt blocks are present in Up and Downsampling layers as well. Deep Supervision is used at each decoder layer, with lower loss weights at lower resolutions. All residuals are \textit{additive} while convolutions are padded to retain tensor sizes. \textbf{(b)} Upsampled Kernel (UpKern) initialization of a pair of MedNeXt architectures with similar configurations ($\theta$) except kernel size ($k_1,k_2$). \textbf{(c)} MedNeXt-L ($5\times5\times5$) leaderboard performance.
    }
    
\end{figure}

\subsection{Fully ConvNeXt 3D Segmentation Architecture}
\label{sec:network}
In prior work, ConvNeXt \cite{liu2022convnet} distilled architectural insights from Vision Transformers \cite{dosovitskiy2020image} and Swin Transformers \cite{liu2021swin} into a convolutional architecture. The ConvNeXt block inherited a number of significant design choices from Transformers, designed to limit computation costs while scaling the network, which demonstrated performance improvements over standard ResNets \cite{he2016deep}. 
In this work, we leverage these strengths by adopting the general design of ConvNeXt as the building block in a 3D-UNet-like \cite{cciccek20163d} macro architecture to obtain the \textbf{MedNeXt}. We extend these blocks to up and downsampling layers as well (Sec. \ref{sec:resampling}), resulting in the first fully ConvNeXt architecture for medical image segmentation. The macro architecture is illustrated in Figure \ref{fig:network_design}.
MedNeXt blocks (similar to ConvNeXt blocks) have 3-layers mirroring a Transformer block and are described for a $C$-channel input as follows: 

\begin{enumerate}
    \item \textbf{Depthwise Convolution Layer:} This layer contains a Depthwise Convolution with kernel size $k\times k \times k$, followed by normalization, with $C$ output channels.
    We use channel-wise GroupNorm \cite{wu2018group} for stability with small batches \cite{roy20222}, instead of the original LayerNorm.
    The depthwise nature of convolutions allow large kernels in this layer to replicate a large attention window of Swin-Transformers, while simultaneously limiting compute and thus delegating the ``heavy lifting" to the Expansion Layer. 
    \item \textbf{Expansion Layer:} Corresponding to a similar design in Transformers, this layer contains an overcomplete Convolution Layer with $CR$ output channels, where $R$ is the expansion ratio, followed by a GELU \cite{hendrycks2016gaussian} activation. Large values of $R$ allow the network to scale \textit{width-wise} while $1\times1\times1$ kernel limits compute. It is important to note that this layer effectively decouples width scaling from receptive field (kernel size) scaling in the previous layer.
    \item \textbf{Compression Layer:} Convolution layer with $1\times1\times1$ kernel and $C$ output channels performing channel-wise compression of the feature maps.
\end{enumerate}

MedNeXt is convolutional and retains the inductive bias inherent to Conv-Nets that allows easier training on sparse medical datasets. Our fully ConvNeXt architecture also enables width (more channels) and receptive field (larger kernels) scaling at both standard and up/downsampling layers. Alongside depth scaling (more layers), we explore these 3 orthogonal types of scaling to design a \textit{compound scalable} MedNeXt for effective medical image segmentation (Sec.\ref{sec:scaling}).

\subsection{Resampling with Residual Inverted Bottlenecks}
\label{sec:resampling}
The original ConvNeXt design utilizes separate downsampling layers which consist of standard strided convolutions. An equivalent upsampling block would be standard strided transposed convolutions. However, this design does not implicitly take advantage of width or kernel-based ConvNeXt scaling while resampling. We improve upon this by extending the Inverted Bottleneck to resampling blocks in MedNeXt. This is done by inserting the strided convolution or transposed convolution in the first \textit{Depthwise Layer} for Downsampling and Upsampling MedNeXt blocks respectively. The corresponding channel reduction or increase is inserted in the last \textit{compression} layer of our MedNeXt $2\times$ Up or Down block design as in Fig. \ref{fig:network_design}. Additionally, to enable easier gradient flow, we add a residual connection with $1\times1 \times 1$ convolution or transposed convolution with \textit{stride} of 2. In doing so, MedNeXt fully leverages the benefits from \textit{Transformer-like} inverted bottlenecks to preserve rich semantic information in lower spatial resolutions in all its components, which should benefit dense medical image segmentation tasks.

\begin{table}[t]
    \centering
    \begin{adjustbox}{width=0.43\textwidth}
    \begin{tabular}{|c|c|c|}
    \hline
    \textbf{Config.} & \textbf{\# Blocks (\textit{B})} & \textbf{Exp. Rat. (\textit{R})}  \\ \hline \hline
    \textbf{S} & \multirow{2}{*}{$B_{all}=2$} & $R_{all}$ = 2 \\ \cline{1-1} \cline{3-3}
    \textbf{B} &  & \multirow{2}{*}{\begin{tabular}[c]{@{}c@{}} $R_1 = R_9 = 2$ \\ $R_2 = R_8 = 3$ \\ $R_{3-7} = 4$ \end{tabular}} \\ \cline{1-2} 
    \textbf{M} & \begin{tabular}[c]{@{}c@{}} $B_1 = B_9 = 3$\\ $B_{2-8}  = 4$\end{tabular} &   \\ \cline{1-3}  
    \textbf{L} & \begin{tabular}[c]{@{}c@{}} $B_1 = B_9 = 3$\\ $B_2 = B_8  = 4$ \\ $B_{3-7}=8$\end{tabular}  & \begin{tabular}[c]{@{}c@{}} $R_1 = R_9 = 3$\\ $R_2 = R_8  = 4$ \\ $R_{3-7}=8$\end{tabular}  \\ \hline
    \end{tabular}    
    \end{adjustbox}
    \begin{adjustbox}{width=0.56\textwidth}
    \rowcolors{3}{}{tablegray}
    \begin{tabular}{|c||cc|cc|}
        \hline
        \multirow{2}{*}{\textbf{Network Variants}} & \multicolumn{2}{c|}{\textbf{BTCV}} & \multicolumn{2}{c|}{\textbf{AMOS22}} \\ \cline{2-5} 
         & \multicolumn{1}{c|}{\textbf{DSC}} & \textbf{SDC} & \multicolumn{1}{c|}{\textbf{DSC}} & \textbf{SDC} \\ \hline \hline
        \textbf{MedNeXt-B Resampling} & \multicolumn{1}{c|}{\textbf{84.01}} & \textbf{86.77} & \multicolumn{1}{c|}{\textbf{89.14}} & \textbf{92.10} \\ \hline
        Standard Up and Downsampling & \multicolumn{1}{c|}{83.13} & 85.64 & \multicolumn{1}{c|}{88.96} & 91.86 \\ \hline \hline
        \textbf{MedNeXt-B ($\mathbf{5\times5\times5})$ + UpKern} & \multicolumn{1}{c|}{\textbf{84.23}} & \textbf{87.06} & \multicolumn{1}{c|}{\textbf{89.38}} & \textbf{92.36} \\ \hline
        MedNeXt-B ($5\times5\times5$) from scratch & \multicolumn{1}{c|}{84.03} & 86.71 & \multicolumn{1}{c|}{89.12} & 92.10 \\ \hline \hline
        \textbf{MedNeXt-B ($\mathbf{5\times5\times5})$ + UpKern} & \multicolumn{1}{c|}{\textbf{84.23}} & \textbf{87.06} & \multicolumn{1}{c|}{\textbf{89.38}} & \textbf{92.36} \\ \hline
        MedNeXt-B ($3\times3\times3$) trained $2\times$ & \multicolumn{1}{c|}{84.00} & 86.85 & \multicolumn{1}{c|}{89.18} & 92.16 \\ \hline \hline
        \textbf{nnUNet} \textit{(non-MedNeXt baseline)} & \multicolumn{1}{c|}{83.56} & 86.07 & \multicolumn{1}{c|}{88.88} & 91.70 \\ \hline
    \end{tabular}
    \end{adjustbox}
    \caption{\textbf{(Left)} MedNeXt configurations from scaling Block Counts ($B$) and Expansion Ratio ($R$) as in Figure \ref{fig:network_design}. \textbf{(Right)} MedNext-B ablations (Sec. \ref{sec:res_abl}) .\label{tab:configs_and_ablation}}
\end{table}  

\subsection{UpKern: Large Kernel Convolutions without Saturation}
\label{sec:upkern}
Large convolution kernels approximate the large attention windows in Transformers, but remain prone to performance saturation. ConvNeXt architectures in classification of natural images, despite the benefit of large datasets such as ImageNet-1k and ImageNet-21k, are seen to saturate at kernels of size $7\times7\times7$ \cite{liu2022convnet}. 
Medical image segmentation tasks have significantly less data and performance saturation can be a problem in large kernel networks. 
To propose a solution, we borrow inspiration from Swin Transformer V2 \cite{liu2022swin} where a large-attention-window network is initialized with another network trained with a smaller attention window. Specifically, Swin Transformers use a bias matrix $\hat{B} \in \mathbb{R}^{(2M-1)\times (2M-1)}$ to store learnt relative positional embeddings, where $M$ is the number of patches in an attention window. On increasing the window size, $M$ increases and necessitates a larger $\hat{B}$. The authors proposed spatially interpolating an existing bias matrix to the larger size as a pretraining step, instead of training from scratch, which demonstrated improved performance. We propose a similar approach but customized to convolutions kernels, as seen in Figure \ref{fig:upkern}, to overcome performance saturation. \textbf{UpKern} allows us to iteratively increase kernel size by initializing a large kernel network with a \textit{compatible} pretrained small kernel network by \textit{trilinearly upsampling} convolutional kernels (represented as tensors) of incompatible size. All other layers with identical tensor sizes (including normalization layers) are initialized by copying the unchanged pretrained weights. This leads to a simple but effective initialization technique for MedNeXt which helps large kernel networks overcome performance saturation in the comparatively limited data scenarios common to medical image segmentation.

\subsection{Compound Scaling of Depth, Width and Receptive Field}
\label{sec:scaling}
\textit{Compound scaling} \cite{tan2019efficientnet} is the idea that simultaneous scaling on multiple levels (depth, width, receptive field, resolution etc) offers benefits beyond that of scaling at one single level. The computational requirements of indefinitely scaling kernel sizes in 3D networks quickly becomes prohibitive and leads us to investigate simultaneous scaling at different levels. Keeping with Figure \ref{fig:network_design}, our scaling is tested for block count ($B$), expansion ratio ($R$) and kernel size ($k$) --  corresponding to depth, width and receptive field size. We use 4 model configurations of the MedNeXt to do so, as detailed in Table \ref{tab:configs_and_ablation} \textbf{(Left)}. The basic functional design (MedNeXt-S) uses number of channels ($C$) as 32, $R=2$ and $B=2$. Further variants increase on just $R$ (MedNeXt-B) or both $R$ and $B$ (MedNeXt-M). The largest 62-MedNext-block architecture uses high values of both $R$ and $B$ (MedNeXt-L) and is used to demonstrate the ability of MedNeXt to be significantly scaled depthwise (even at standard kernel sizes). We further explore large kernel sizes and experiment with $k=\{3, 5\}$ for each configuration, to maximize performance via \textit{compound scaling} of the MedNeXt architecture.

\section{Experimental Design}
\subsection{Configurations, Implementation and Baselines}
We use PyTorch \cite{paszke2019pytorch} for implementing our framework. We experiment with 4 configurations of the MedNeXt with 2 kernel sizes as detailed in Section \ref{sec:scaling}. The GPU memory requirements of scaling are limited via -- 1) Mixed precision training with \texttt{PyTorch AMP}, 2) Gradient Checkpointing. \cite{chen2016training}. 
Our experimental framework uses the nnUNet \cite{isensee2021nnu} as a backbone - where the training schedule (epochs=1000, batches per epoch=250), inference (50\% patch overlap) and data augmentation remain unchanged. All networks, except nnUNet, are trained with AdamW \cite{loshchilov2017decoupled} as optimizer. The data is resampled to $1.0$ mm isotropic spacing during training and inference (with results on original spacing), using input patch size of $128\times 128 \times 128$ and $512 \times 512$, and batch size 2 and 14, for 3D and 2D networks respectively. The learning rate for all MedNeXt models is $0.001$, except kernel:5 in KiTS19, which uses $0.0001$ for stability. For baselines, all Swin models and 3D-UX-Net use $0.0025$, while ViT models use $0.0001$. We use Dice Similarity Coefficient (DSC) and Surface Dice Similarity (SDC) at 1.0mm tolerance for volumetric and surface accuracy. 5-fold cross-validation (CV) mean performance for supervised training using 80:20 splits for all models are reported. We also provide test set DSC scores for a 5-fold ensemble of MedNeXt-L (kernel: $5\times5\times5$) without postprocessing. Our extensive baselines consist of a high-performing convolutional network (nnUNet \cite{isensee2021nnu}), 4 convolution-transformer hybrid networks with transformers in the encoder (UNETR \cite{hatamizadeh2022unetr}, SwinUNETR \cite{hatamizadeh2022swin}) and in intermediate layers (TransBTS \cite{wang2021transbts}, TransUNet \cite{chen2021transunet}), a fully transformer network (nnFormer \cite{zhou2021nnformer}) as well as a partially ConvNeXt network (3D-UX-Net \cite{lee20223d}). TransUNet is a 2D network while the rest are 3D networks. The uniform framework provides a common testbed for all networks, without incentivizing one over the other on aspects of patch size, spacing, augmentations, training and evaluation.

\subsection{Datasets}
We use 4 popular tasks, encompassing organ as well as tumor segmentation tasks, to comprehensively demonstrate the benefits of the MedNeXt architecture -- 1) Beyond-the-Cranial-Vault (BTCV) Abdominal CT Organ Segmentation \cite{landman2015miccai}, 2) AMOS22 Abdominal CT Organ Segmentation \cite{ji2022amos} 3) Kidney Tumor Segmentation Challenge 2019 Dataset (KiTS19) \cite{heller2020state}, 4) Brain Tumor Segmentation Challenge 2021 (BraTS21) \cite{baid2021rsna}. BTCV, AMOS22 and KiTS19 datasets contain 30, 200 and 210 CT volumes with 13, 15 and 2 classes respectively, while the BraTS21 dataset contains 1251 MRI volumes with 3 classes. This diversity shows the effectiveness of our methods across imaging modalities and training set sizes.

\section{Results and Discussion}

\begin{table}[t]
\begin{adjustbox}{width=\textwidth}
\begin{tabular}{|r|c|c|c|c|c|c|c|c|c||c|c|}
\hline
\multicolumn{1}{|c|}{\multirow{2}{*}{\textbf{Networks}}} & \multirow{2}{*}{\textbf{Cat.}} & \multicolumn{2}{c|}{\textbf{BTCV}} & \multicolumn{2}{c|}{\textbf{AMOS22}} & \multicolumn{2}{c|}{\textbf{KiTS19}} & \multicolumn{2}{c||}{\textbf{BraTS21}} & \multicolumn{2}{c|}{\textbf{AVG}} \\ \cline{3-12} 
\multicolumn{1}{|c|}{} & & \multicolumn{1}{c|}{\textbf{~DSC ~}} & \multicolumn{1}{c|}{\textbf{~SDC ~}} & \multicolumn{1}{c|}{\textbf{~DSC ~}} & \multicolumn{1}{c|}{\textbf{~SDC ~}} & \multicolumn{1}{c|}{\textbf{~DSC ~}} & \multicolumn{1}{c|}{\textbf{~SDC ~}} & \multicolumn{1}{c|}{\textbf{~DSC ~}} & \multicolumn{1}{c||}{\textbf{~SDC ~}} & \multicolumn{1}{c|}{\textbf{~DSC ~}} & \multicolumn{1}{c|}{\textbf{~SDC ~}} \\ \hline \hline
nnUNet & \multirow{7}{*}{\rotatebox{90}{\textit{Baselines}}} & 83.56 & 86.07 & 88.88 & 91.70 & 89.88 & 86.88 & 91.23 & 90.46 & 88.39 & 88.78 \\
UNETR & & 75.06 & 75.00 & 81.98 & 82.65 & 84.10 & 78.05 & 89.65 & 88.28 & 82.36 & 81.00 \\
TransUNet & & 76.72 & 76.64 & 85.05 & 86.52 & 80.82 & 72.90 & 89.17 & 87.78 & 82.94 & 80.96 \\
TransBTS & & 82.35 & 84.33 & 86.52 & 88.84 & 87.03 & 83.53 & 90.66 & 89.71 & 86.64 & 86.60 \\
nnFormer & & 80.76 & 82.37 & 84.20 & 86.38 & 89.09 & 85.08 & 90.42 & 89.83 & 86.12 & 85.92 \\
SwinUNETR & & 80.95 & 82.43 & 86.83 & 89.23 & 87.36 & 83.09 & 90.48 & 89.56 & 86.41 & 86.08 \\
3D-UX-Net & & 80.76 & 82.30 & 87.28 & 89.74 & 88.39 & 84.03 & 90.63 & 89.63 & 86.77 & 86.43 \\ \hline \hline
MedNeXt-S & \multirow{4}{*}{\rotatebox{90}{\textit{kernel: 3}}} & \Colorone{83.90} & \Colorone{86.60} & \Colorone{89.03} & \Colorone{91.97} & \Colorone{90.45} & \Colorone{87.80} & \Colorone{91.27} & \Colorone{90.46} & \Colorone{88.66} & \Colorone{89.21} \\
MedNeXt-B & & \Colorone{84.01} & \Colorone{86.77} & \Colorone{89.14} & \Colorone{92.10} & \Colorone{\textbf{91.02}} & \Colorone{\textbf{88.24}} & \Colorone{91.30} & \Colorone{90.51} & \Colorone{88.87} & \Colorone{89.41} \\
MedNeXt-M & & \Colorone{84.31} & \Colorone{87.34} & \Colorone{89.27} & \Colorone{92.28} & \Colorone{90.78} & \Colorone{88.22} & \Colorone{\textbf{91.57}} & \Colorone{90.78} & \Colorone{88.98} & \Colorone{89.66} \\
MedNeXt-L & & \Colorone{\textbf{84.57}} & \Colorone{\textbf{87.54}} & \Colorone{\textbf{89.58}} & \Colorone{\textbf{92.62}} & \Colorone{90.61} & \Colorone{88.08} & \Colorone{\textbf{91.57}} & \Colorone{\textbf{90.81}} & \Colorone{\textbf{89.08}} & \Colorone{\textbf{89.76}} \\ \hline
MedNeXt-S & \multirow{4}{*}{\rotatebox{90}{\textit{kernel: 5}}} & \Colortwo{83.92} & \Colortwo{86.80} & \Colortwo{89.27} & \Colortwo{92.26} & \Colorone{90.08} & \Colorone{87.04} & \Colortwo{91.40} & \Colortwo{90.57} & \Colortwo{88.67} & \Colorone{89.17} \\
MedNeXt-B & & \Colortwo{84.23} & \Colortwo{87.06} & \Colortwo{89.38} & \Colortwo{92.36} & \Colorone{90.30} & \Colorone{87.40} & \Colortwo{91.48} & \Colortwo{90.70} & \Colorone{88.85} & \Colorone{89.38} \\
MedNeXt-M & & \Colortwo{84.41} & \Colortwo{87.48} & \Colortwo{89.58} & \Colortwo{92.65} & \Colortwo{\textbf{90.87}} & \Colorone{\textbf{88.15}} & \Colorone{\textbf{91.49}} & \Colorone{90.67} & \Colortwo{89.09} & \Colortwo{89.74} \\
MedNeXt-L & & \Colortwo{\textbf{84.82}} & \Colortwo{\textbf{87.85}} & \Colortwo{\textbf{89.87}} & \Colortwo{\textbf{92.95}} & \Colortwo{90.71} & \Colorone{87.85} & \Colorone{91.46} & \Colorone{\textbf{90.73}} & \Colortwo{\textbf{89.22}} & \Colortwo{\textbf{89.85}} \\ \hline
\end{tabular}
\end{adjustbox}
\caption{5-fold CV results of MedNeXt at kernel sizes: $\{3, 5\}$ outperforming 7 baselines -- consisting of convolutional, transformer and large kernel networks. \\ 
\Colorone{val} (bold): better than or equal to ($\geq$) top baseline\\
\Colortwo{val} (underline): better than ($>$) \textit{kernel: $3$} of same configuration\label{tab:results}
}
\end{table}

\subsection{Performance ablation of architectural improvements}
\label{sec:res_abl}
We ablate the MedNeXt-B configuration on AMOS22 and BTCV datasets to highlight the efficacy of our improvements and demonstrate that a \textit{vanilla} ConvNeXt is unable to compete with existing segmentation baselines such as nnUNet. The following are observed in ablation tests in Table \ref{tab:configs_and_ablation} \textbf{(Right)} -- 
\begin{enumerate}
    \item Residual Inverted Bottlenecks, specifically in Up and Downsampling layers, \textit{functionally enables} MedNeXt (MedNeXt-B Resampling vs Standard Resampling) for medical image segmentation. In contrast, absence of these modified blocks lead to \textbf{considerably worse} performance. This is possibly owing to preservation of semantic richness in feature maps while resampling. 
    \item Training large kernel networks for medical image segmentation is a non-trivial task, with large kernel MedNeXts trained from scratch failing to perform in seen in MedNeXt-B (UpKern vs From Scratch). UpKern \textit{improves performance} in kernel $5\times5\times5$ on both BTCV and AMOS22, whereas large kernel performance is \textbf{indistinguishable} from small kernels \textit{without} it. 
    \item The performance boost in large kernels is seen to be due to the combination of UpKern with a larger kernel and not merely a longer \textit{effective} training schedule (Upkern vs Trained $2\times$), as a trained MedNeXt-B with kernel $3\times3\times3$ retrained again is \textbf{unable to match} its large kernel counterpart.
\end{enumerate} 

This highlights that the MedNeXt modifications successfully translate the ConvNeXt architecture to medical image segmentation. We further establish the performance of the MedNeXt architecture against our baselines -- comprising of convolutional, transformer-based and large kernel baselines -- on all 4 datasets. We discuss the effectiveness of the MedNeXt on multiple levels.

\subsection{Performance comparison to baselines}
There are 2 levels at which MedNeXt successfully overcomes existing baselines - 5 fold CV and public testset performance. In 5-fold CV scores in Table \ref{tab:results}, MedNeXt, with $3\times3\times3$ kernels, takes advantage of depth and width scaling to provide state-of-the-art segmentation performance against \textbf{every baseline on all 4 datasets} with no additional training data. MedNeXt-L outperforms or is competitive with smaller variants despite task heterogeneity (brain and kidney tumors, organs), modality (CT, MRI) and training set size (BTCV: 18 samples vs BraTS21: 1000 samples), establishing itself as a powerful alternative to established methods such as nnUNet.
With UpKern and $5\times5\times5$ kernels, MedNeXt takes advantage of full compound scaling to \textbf{improve further} on its own small kernel networks, comprehensively on organ segmentation (BTCV, AMOS22) and in a more limited fashion on tumor segmentation (KiTS19, BraTS21). 

Furthermore, in leaderboard scores on official testsets (Fig. \ref{fig:testset_scores}), 5-fold ensembles for MedNeXt-L (kernel: $5\times5\times5$) and nnUNet, its strongest competitor are compared -- \textbf{1) \underline{BTCV}:} MedNeXt beats nnUNet and, to the best of our knowledge, is one of the leading methods with \textit{only supervised training} and \textit{no extra training data} (DSC: 88.76, HD95: 15.34), \textbf{2) \underline{AMOS22}:} MedNeXt not only surpasses nnUNet, but is also \textbf{Rank 1} (date: 09.03.23) currently on the leaderboard (DSC: 91.77, NSD: 84.00), \textbf{3) \underline{KITS19}:} MedNeXt exceeds nnUNet performance (DSC: 91.02), \textbf{4) \underline{BraTS21}:} MedNeXt surpasses nnUNet in both volumetric and surface accuracy (DSC: 88.01, HD95: 10.69). MedNeXt attributes its performance solely to its architecture without leveraging techniques like transfer learning (3D-UX-Net) or repeated 5-fold ensembling (UNETR, SwinUNETR), thus establishing itself as the state-of-the-art for medical image segmentation.

\section{Conclusion}
In comparison to natural image analysis, medical image segmentation lacks architectures that benefit from scaling networks due to inherent domain challenges such as limited training data. In this work, MedNeXt is presented as a scalable \textit{Transformer-inspired} fully-ConvNeXt 3D segmentation architecture customized for high performance on limited medical image datasets. We demonstrate MedNeXt's state-of-the-art performance across 4 challenging tasks against 7 strong baselines. Additionally, similar to ConvNeXt for natural images \cite{liu2022convnet}, we offer the \textit{compound scalable} MedNeXt design as an effective modernization of standard convolution blocks for building deep networks for medical image segmentation.

\bibliographystyle{splncs04}
\bibliography{references.bib}

\begin{thebibliography}{10}
\providecommand{\url}[1]{\texttt{#1}}
\providecommand{\urlprefix}{URL }
\providecommand{\doi}[1]{https://doi.org/#1}

\bibitem{baid2021rsna}
Baid, U., Ghodasara, S., Mohan, S., Bilello, M., Calabrese, E., Colak, E.,
  Farahani, K., Kalpathy-Cramer, J., Kitamura, F.C., Pati, S., et~al.: The
  rsna-asnr-miccai brats 2021 benchmark on brain tumor segmentation and
  radiogenomic classification. arXiv preprint arXiv:2107.02314  (2021)

\bibitem{cao2021swin}
Cao, H., Wang, Y., Chen, J., Jiang, D., Zhang, X., Tian, Q., Wang, M.:
  Swin-unet: Unet-like pure transformer for medical image segmentation. arXiv
  preprint arXiv:2105.05537  (2021)

\bibitem{chen2021transunet}
Chen, J., Lu, Y., Yu, Q., Luo, X., Adeli, E., Wang, Y., Lu, L., Yuille, A.L.,
  Zhou, Y.: Transunet: Transformers make strong encoders for medical image
  segmentation. arXiv preprint arXiv:2102.04306  (2021)

\bibitem{chen2016training}
Chen, T., Xu, B., Zhang, C., Guestrin, C.: Training deep nets with sublinear
  memory cost. arXiv preprint arXiv:1604.06174  (2016)

\bibitem{cciccek20163d}
{\c{C}}i{\c{c}}ek, {\"O}., Abdulkadir, A., Lienkamp, S.S., Brox, T.,
  Ronneberger, O.: 3d u-net: learning dense volumetric segmentation from sparse
  annotation. In: International conference on medical image computing and
  computer-assisted intervention. pp. 424--432. Springer (2016)

\bibitem{deng2009imagenet}
Deng, J., Dong, W., Socher, R., Li, L.J., Li, K., Fei-Fei, L.: Imagenet: A
  large-scale hierarchical image database. In: 2009 IEEE conference on computer
  vision and pattern recognition. pp. 248--255. Ieee (2009)

\bibitem{dosovitskiy2020image}
Dosovitskiy, A., Beyer, L., Kolesnikov, A., Weissenborn, D., Zhai, X.,
  Unterthiner, T., Dehghani, M., Minderer, M., Heigold, G., Gelly, S., et~al.:
  An image is worth 16x16 words: Transformers for image recognition at scale.
  arXiv preprint arXiv:2010.11929  (2020)

\bibitem{hatamizadeh2022swin}
Hatamizadeh, A., Nath, V., Tang, Y., Yang, D., Roth, H.R., Xu, D.: Swin unetr:
  Swin transformers for semantic segmentation of brain tumors in mri images.
  In: International MICCAI Brainlesion Workshop. pp. 272--284. Springer (2022)

\bibitem{hatamizadeh2022unetr}
Hatamizadeh, A., Tang, Y., Nath, V., Yang, D., Myronenko, A., Landman, B.,
  Roth, H.R., Xu, D.: Unetr: Transformers for 3d medical image segmentation.
  In: Proceedings of the IEEE/CVF Winter Conference on Applications of Computer
  Vision. pp. 574--584 (2022)

\bibitem{he2016deep}
He, K., Zhang, X., Ren, S., Sun, J.: Deep residual learning for image
  recognition. In: Proceedings of the IEEE conference on computer vision and
  pattern recognition. pp. 770--778 (2016)

\bibitem{heller2020state}
Heller, N., Isensee, F., Maier-Hein, K.H., Hou, X., Xie, C., Li, F., Nan, Y.,
  Mu, G., Lin, Z., Han, M., et~al.: The state of the art in kidney and kidney
  tumor segmentation in contrast-enhanced ct imaging: Results of the kits19
  challenge. Medical Image Analysis p. 101821 (2020)

\bibitem{hendrycks2016gaussian}
Hendrycks, D., Gimpel, K.: Gaussian error linear units (gelus). arXiv preprint
  arXiv:1606.08415  (2016)

\bibitem{isensee2021nnu}
Isensee, F., Jaeger, P.F., Kohl, S.A., Petersen, J., Maier-Hein, K.H.: nnu-net:
  a self-configuring method for deep learning-based biomedical image
  segmentation. Nature methods  \textbf{18}(2),  203--211 (2021)

\bibitem{ji2022amos}
Ji, Y., Bai, H., Yang, J., Ge, C., Zhu, Y., Zhang, R., Li, Z., Zhang, L., Ma,
  W., Wan, X., et~al.: Amos: A large-scale abdominal multi-organ benchmark for
  versatile medical image segmentation. arXiv preprint arXiv:2206.08023  (2022)

\bibitem{karimi2021convolution}
Karimi, D., Vasylechko, S.D., Gholipour, A.: Convolution-free medical image
  segmentation using transformers. In: Medical Image Computing and Computer
  Assisted Intervention--MICCAI 2021: 24th International Conference,
  Strasbourg, France, Proceedings, Part I 24. pp. 78--88. Springer (2021)

\bibitem{landman2015miccai}
Landman, B., Xu, Z., Igelsias, J., Styner, M., Langerak, T., Klein, A.: Miccai
  multi-atlas labeling beyond the cranial vault--workshop and challenge. In:
  Proc. MICCAI Multi-Atlas Labeling Beyond Cranial Vault Challenge. vol.~5,
  p.~12 (2015)

\bibitem{lee20223d}
Lee, H.H., Bao, S., Huo, Y., Landman, B.A.: 3d ux-net: A large kernel
  volumetric convnet modernizing hierarchical transformer for medical image
  segmentation. arXiv preprint arXiv:2209.15076  (2022)

\bibitem{li2023large}
Li, H., Nan, Y., Del~Ser, J., Yang, G.: Large-kernel attention for 3d medical
  image segmentation. Cognitive Computation pp. 1--15 (2023)

\bibitem{litjens2017survey}
Litjens, G., Kooi, T., Bejnordi, B.E., Setio, A.A.A., Ciompi, F., Ghafoorian,
  M., Van Der~Laak, J.A., Van~Ginneken, B., S{\'a}nchez, C.I.: A survey on deep
  learning in medical image analysis. Medical image analysis  \textbf{42},
  60--88 (2017)

\bibitem{liu2022swin}
Liu, Z., Hu, H., Lin, Y., Yao, Z., Xie, Z., Wei, Y., et~al.: Swin transformer
  v2: Scaling up capacity and resolution. In: Proceedings of the IEEE/CVF
  Conference on Computer Vision and Pattern Recognition. pp. 12009--12019
  (2022)

\bibitem{liu2021swin}
Liu, Z., Lin, Y., Cao, Y., Hu, H., Wei, Y., Zhang, Z., et~al.: Swin
  transformer: Hierarchical vision transformer using shifted windows. In:
  Proceedings of the IEEE/CVF International Conference on Computer Vision. pp.
  10012--10022 (2021)

\bibitem{liu2022convnet}
Liu, Z., Mao, H., Wu, C.Y., Feichtenhofer, C., Darrell, T., Xie, S.: A convnet
  for the 2020s. In: Proceedings of the IEEE/CVF Conference on Computer Vision
  and Pattern Recognition. pp. 11976--11986 (2022)

\bibitem{loshchilov2017decoupled}
Loshchilov, I., Hutter, F.: Decoupled weight decay regularization. arXiv
  preprint arXiv:1711.05101  (2017)

\bibitem{paszke2019pytorch}
Paszke, A., Gross, S., Massa, F., Lerer, A., Bradbury, J., Chanan, G., Killeen,
  T., et~al.: Pytorch: An imperative style, high-performance deep learning
  library. Advances in neural information processing systems  \textbf{32}
  (2019)

\bibitem{peiris2022robust}
Peiris, H., Hayat, M., Chen, Z., et~al.: A robust volumetric transformer for
  accurate 3d tumor segmentation. In: International Conference on Medical Image
  Computing and Computer-Assisted Intervention. pp. 162--172. Springer (2022)

\bibitem{ridnik2021imagenet}
Ridnik, T., Ben-Baruch, E., Noy, A., Zelnik-Manor, L.: Imagenet-21k pretraining
  for the masses. arXiv preprint arXiv:2104.10972  (2021)

\bibitem{roy20222}
Roy, S., K{\"u}gler, D., Reuter, M.: Are 2.5 d approaches superior to 3d deep
  networks in whole brain segmentation? In: International Conference on Medical
  Imaging with Deep Learning. pp. 988--1004. PMLR (2022)

\bibitem{simonyan2014very}
Simonyan, K., Zisserman, A.: Very deep convolutional networks for large-scale
  image recognition. arXiv preprint arXiv:1409.1556  (2014)

\bibitem{tan2019efficientnet}
Tan, M., Le, Q.: Efficientnet: Rethinking model scaling for convolutional
  neural networks. In: International conference on machine learning. pp.
  6105--6114. PMLR (2019)

\bibitem{vaswani2017attention}
Vaswani, A., Shazeer, N., Parmar, N., Uszkoreit, J., Jones, L., Gomez, A.N.,
  Kaiser, {\L}., Polosukhin, I.: Attention is all you need. Advances in neural
  information processing systems  \textbf{30} (2017)

\bibitem{wang2021transbts}
Wang, W., Chen, C., Ding, M., Yu, H., et~al.: Transbts: Multimodal brain tumor
  segmentation using transformer. In: International Conference on Medical Image
  Computing and Computer-Assisted Intervention. pp. 109--119. Springer (2021)

\bibitem{wu2018group}
Wu, Y., He, K.: Group normalization. In: Proceedings of the European conference
  on computer vision (ECCV). pp. 3--19 (2018)

\bibitem{xie2021cotr}
Xie, Y., Zhang, J., Shen, C., Xia, Y.: Cotr: Efficiently bridging cnn and
  transformer for 3d medical image segmentation. In: International conference
  on medical image computing and computer-assisted intervention. pp. 171--180.
  Springer (2021)

\bibitem{zhou2021nnformer}
Zhou, H.Y., Guo, J., Zhang, Y., Yu, L., Wang, L., Yu, Y.: nnformer: Interleaved
  transformer for volumetric segmentation. arXiv preprint arXiv:2109.03201
  (2021)

\end{thebibliography}

\newpage
\section{Supplementary Material}

\subsection{UpKern: Algorithm}
UpKern Initialization can be condensed into 20 lines of \texttt{PyTorch} pseudo-code. While a 3D architecture is used in this work, the technique is general enough to be translated to 2D by changing `trilinear' interpolation to `bilinear'.
\begin{algorithm}[H] \small
    \centering
    \caption{PyTorch code in 20 lines for Upsampled Kernel Initialization (UpKern) for MedNeXt with large kernels ($\geq 5$) }
    \label{alg:cap}
\begin{lstlisting}[language=Python]
import torch.functional as F

def upkern_init_load_weights(network, pretrained_net):
    
    pretrn_dct = pretrained_net.state_dict()
    model_dct = network.state_dict()
    for k in model_dct.keys():
        # Only common keys and core algorithm in Demo code
        if k in model_dct.keys() and k in pretrn_dct.keys():  
            inc1, outc1, *spt_dims1 = model_dct[k].shape
            inc2, outc2, *spt_dims2 = pretrn_dct[k].shape
            if spt_dims1 == spt_dims2:     # standard init
                model_dct[k] = pretrn_dct[k]
            else: # Upsampled kernel init
                model_dct[k] = F.interpolate(     
                                    pretrn_dct[k],
                                    size=spt_dims1,
                                    mode='trilinear')
    network.load_state_dict(model_dct)
    return network
\end{lstlisting}
    \label{alg:upkern}
\end{algorithm}

\subsection{Detailed Model Configuration}
\begin{table}[H]
\centering
\label{tab:archs}
\begin{adjustbox}{width=0.9\textwidth, height=1.3cm}
\begin{tabular}{|c|c|c|c|c|c|c|}
\hline
\textbf{Configs} & \textbf{\# Blocks (\textit{B})} & \textbf{Exp. Ratio (\textit{R})} & \textbf{\textit{C}} & \textbf{Params (Mi)} & \textbf{GFLOPs} & \textbf{secs/Train Epoch} \\ \hline
\textbf{S} & \multirow{2}{*}{$B_{all}=2$} & $R_{all}$ = 2 & \multirow{7}{*}{32} & 5.6, 5.9 & 130, 169 & 117, 364 \\ 
\cline{1-1} \cline{3-3} \cline{5-7} 
\textbf{B} &  & \multirow{2}{*}{\begin{tabular}[c]{@{}c@{}} $R_1 = R_9 = 2$ \\ $R_2 = R_8 = 3$ \\ $R_{3-7} = 4$ \end{tabular}} &  & 10.5, 11.0 & 170, 208 & 120, 381 \\ 
\cline{1-2} \cline{5-7} 
\textbf{M} & \begin{tabular}[c]{@{}c@{}} $B_1 = B_9 = 3$\\ $B_{2-8}  = 4$\end{tabular} &  &  & 17.6, 18.3 & 248, 308 & 213, 668 \\ 
\cline{1-3}  \cline{5-7}
\textbf{L} & \begin{tabular}[c]{@{}c@{}} $B_1 = B_9 = 3$\\ $B_2 = B_8  = 4$ \\ $B_{3-7}=8$\end{tabular}  & \begin{tabular}[c]{@{}c@{}} $R_1 = R_9 = 3$\\ $R_2 = R_8  = 4$ \\ $R_{3-7}=8$\end{tabular} &  & 61.8, 63.0 & 500, 564 &  267, 735 \\ \hline
\end{tabular}
\end{adjustbox}
\caption{Detailed configurations, GLOPs and training time at kernel=\{3, 5\} of MedNeXt architectures (left) in this work. GFLOPs is calculated for a single $128 \times 128 \times 128$ patch. Training time is provided of a single epoch with batch size = 2 and 250 batches.}
\end{table}

\section{Tabulation of Results on 4 Test Sets}

\begin{table}[]
    \centering
    \begin{tabular}{c|c|c}
        Technique & Dataset & DSC \\ \hline
        \multirow{4}{*}{MedNeXt} & BTCV & 88.76 \\ 
        & AMOS22 & 91.77 \\ 
        & KiTS19 & 91.02 \\ 
        & BraTS21 & 88.01 \\ \hline
    \end{tabular}
    \caption{Tabulation of results on public validation sets from Figure \ref{fig:testset_scores} of the 4 datasets -- BTCV, AMOS22, KiTS19 and BraTS21 -- used in this work. A single 5-fold ensemble of MedNeXt-L using $5 \times 5 \times 5$ kernels, initialized using UpKern were used for all datasets. The \texttt{3d\_fullres} mode of the nnUNet pipeline was used and the \texttt{cascade} mode was not utilized.}
    \label{tab:my_label}
\end{table}

\end{document}